\documentclass[11pt]{article}
\usepackage{graphicx}
\usepackage{amssymb}
\usepackage{amsmath}
\usepackage{graphicx}
\usepackage{amstext}
\usepackage{leftidx}
\usepackage{esint}
\usepackage[utf8]{inputenc}
\usepackage{color}
\usepackage{cite}
\usepackage{hyperref}
\usepackage{multido}

\newcommand\ringring[1]{%
  {
   \mathop{\kern0pt #1}\limits^{
     \vbox to-1.85ex{
       \kern-2ex 
       \hbox to 0pt{\hss\normalfont\kern.1em \r{}\kern-.45em \r{}\hss}%
       \vss 
     }
   }
  }
} 

\makeatletter
\ams@newcommand{\vardot}[2]{%
  {\mathop{#2\kern0pt}\limits^{\vbox to-1.4\ex@{\kern-\tw@\ex@
   \hbox{\normalfont\multido{}{#1}{.}}\vss}}}}
\makeatother
\newcommand{\CC}{\Lambda}

\newcommand{\rv}{\rho_{\rm vac}}

\newcommand{\rvo}{\rho^0_{\rm vac}}

\newcommand{\rmr}{\rho_m}

\newcommand{\rX}{\rho_X}

\newcommand{\wX}{\omega_X}
\newcommand{\wY}{\omega_Y}

\newcommand{\CCS}{\CC_s{\rm CDM}}
\newcommand{\wXCDM}{$w${\rm XCDM}\,}

\newcommand{\nueff}{\nu_{\rm eff}}

\newcommand{\mpl}{m_{\rm Pl}}

\newcommand{\be}{\begin{equation}}
\newcommand{\ee}{\end{equation}}

\begin{document}

\hyphenation{theo-re-ti-cal gra-vi-ta-tio-nal theo-re-ti-cally mo-dels cos-mo-lo-gi-cal cor-res-pon-ding}

\begin{center}
\vskip 2mm
{\bf \Large  Composite Dark Energy and the Cosmological Tensions}
\vskip 2mm

 \vskip 8mm

\textbf{Adri\`a G\'omez-Valent, Joan Sol\`a Peracaula}

\vskip 0.4cm

Departament de F\'isica Qu\`antica i Astrof\'isica, \\
and   Institute of Cosmos Sciences,\\ Universitat de Barcelona,
Av. Diagonal 647, E-08028 Barcelona, Catalonia, Spain

\vskip 0.5cm

E-mails:  agomezvalent@icc.ub.edu, sola@fqa.ub.edu

 \vskip2mm

\end{center}
\vskip 15mm

\begin{quotation}
\noindent {\large\it \underline{Abstract}}.
The standard cosmological model currently in force, aka $\Lambda$CDM, has been plagued with a variety of phenomenological glitches or tensions in the last decade or so, which puts it against the wall.  At the core of the  $\Lambda$CDM we have a rigid cosmological term, $\Lambda$,  for the entire cosmic history. This feature is  unnatural and even inconsistent in the context of  fundamental physics.  Recently, the results from the DESI collaboration suggested the possibility that dark energy (DE) should be dynamical rather than just a cosmological constant. Using a generic  $w_0w_a$CDM parameterization, DESI reported signs of  quintessence behavior at $2.5-3.9\sigma$ c.l. by combining their BAO data with CMB and different SNIa samples. However, to alleviate the tensions the DE needs more features.  In the proposed
$w$XCDM model\,[45],  the DE is actually  a composite  cosmic fluid  with two components $(X,Y)$ acting  sequentially: first  $X$ (above a transition redshift $z_t$)  and second  $Y$ (below $z_t$). Fitting the model to the data, we find that the late component $Y$ behaves as quintessence, like DESI. However, to cure the $H_0$ and growth tensions,  $X$  must  behave as  `phantom matter' (PM), which in contrast to phantom DE provides positive pressure at the expense of negative energy density. The PM behavior actually appears in stringy versions of the running vacuum model (RVM).  Using the SNIa (considering  separately Pantheon$+$ and DESY5), cosmic chronometers, transversal BAO, LSS data, and the full CMB likelihood from Planck 2018, we find that both tensions can be completely cut down. We also compare the  $w$XCDM  with our own results using the standard $w$CDM and  $w_0w_a$CDM parameterizations of the DE.  In all cases, model $w$XCDM performs much better. Finally, we have repeated our analysis with BAO 3D data (replacing BAO 2D), and we still find that the main dynamical DE models (including composite ones) provide a much better fit quality compared to $\Lambda$CDM. The growth tension is alleviated again, but in contrast, the $H_0$-tension remains significant, which is most likely reminiscent of the internal conflict in the BAO sector.


\end{quotation}

\newpage


\newpage
\section{Introduction}

Despite the standard or concordance model of cosmology, aka $\CC$CDM\,\cite{peebles:1993},  has proven to be a rather successful  theoretical framework for the description of the Universe's evolution as a whole, in the last few decades the model  has led to a number of  pitfalls which shed  doubts about its phenomenological  viability, not to mention other (much deeper) theoretical problems.  A core ingredient of the model is  the cosmological constant (CC), $\CC$,   introduced by A. Einstein more than a century  ago\,\cite{Einstein:1917}.  The CC is traditionally connected to the vacuum energy density (VED) through $\rv=\CC/(8\pi G_N)$,  with $G_N$  Newton's gravitational constant.  It was later observed that this connection may generate a big theoretical conundrum, the so-called  `cosmological constant problem' (CCP)\,\cite{Zeldovich:1967}. Basically, it stems from the inordinately large prediction on the value of the VED (and so of $\CC$)  made by  quantum field theory (QFT) and in particular by the standard model of  elementary particles. Such a weird prediction amounts indeed to a value of $\rv$ which is at least 55 orders of magnitude  bigger than the current critical density of the Universe\cite{Weinberg:1988cp,Peebles:2002gy,Padmanabhan:2002ji}. Whether this interpretation of the CCP is correct or not is a matter of debate. In the meantime,  recent theoretical developments in the context of the running vacuum model (RVM)\cite{SolaPeracaula:2022hpd,SolaPeracaula:2024nsz,Sola:2013gha,Sola:2015rra} have provided a new QFT perspective for  tackling the CCP and the tensions themselves. In the new context, the VED, and hence also $\CC$, are free from the very  large (quartic mass)  $\sim m^4$ effects associated with the vacuum fluctuations of the quantized matter fields, which are at the root of the need for extreme fine tuning  in the CCP context\,\cite{Moreno-Pulido:2020anb,Moreno-Pulido:2022phq,Moreno-Pulido:2022upl,Moreno-Pulido:2023ryo}.  Moreover, in this  framework  the VED  evolves smoothly with the cosmological expansion, i.e. $\rv=\rv(H)$,  rather than retaining a rigid value $\rv=$const.  for the entire cosmic history, specifically the evolution is like  $\delta\rv(H)\sim \nu H^2$ in the current universe, where $|\nu|\ll1$ is a calculable coefficient in QFT.  Such a dynamics of the VED  has been  substantiated in the context of QFT in curved spacetime in the previously mentioned works as well as in the framework of low-energy effective string theory \cite{Mavromatos:2020kzj,Mavromatos:2021urx,Gomez-Valent:2023hov,Dorlis:2024yqw,Dorlis:2024uei,Basilakos:2019acj}. Very recently, the evolution law $\delta\rv(H)\sim \nu H^2$ has  also been highlighted  in lattice quantum gravity studies\,\cite{Dai:2024vjc}.   With these developments at hand, the CCP has gained a new perspective\,\footnote{An informal introduction to the Cosmological Constant Problem can be found in  \cite{JSP:Cosmoverse2024}. For a more formal approach, see the reviews \cite{SolaPeracaula:2022hpd,SolaPeracaula:2024nsz,Sola:2013gha,Sola:2015rra}, which are framed  along the lines of the current work.}. Intriguingly enough,  this year's release of the measurements by the  Dark Energy Spectroscopic Instrument (DESI), when combined with CMB data from Planck and different SNIa samples, suggest tantalizing evidence of dynamical DE\cite{DESI:2024mwx,DESI:2024hhd}, which is in accordance with the mentioned theoretical works.  The possibility  of dynamical DE was strongly emphasized a few years ago from devoted studies involving a large set of cosmological data\cite{Sola:2015wwa,Sola:2016jky,SolaPeracaula:2016qlq,SolaPeracaula:2017esw} using the framework of the RVM.   Other analyses from different perspectives reached similar conclusions around that time\,\cite{Sahni:2014ooa,Gomez-Valent:2014rxa,Zhao:2017cud,SolaPeracaula:2018wwm}.

The phenomenological pitfalls of the $\CC$CDM  referred to above are concerned with the persisting tensions involved in the measurement of the current Hubble parameter $H_0\equiv 100 h$ km/s/Mpc ($h\simeq 0.7)$ and the growth of large scale structures (LSS), see \cite{DiValentino:2020zio,DiValentino:2020vvd} for summarized explanations about each of these tensions, and \cite{Perivolaropoulos:2021jda,Abdalla:2022yfr,Dainotti:2023yrk} for comprehensive reviews and lists of references.  The first sort of tension leads to  a  serious mismatch  between the  CMB observations when analyzed under the assumption of $\CC$CDM cosmology, and the local (distance ladder) measurements of the Hubble parameter today. It amounts to a severe inconsistency of $\sim 5 \sigma$ CL between the two kinds of  determinations of the $H_0$ value. The second kind of tension points to an overproduction of large scale structure at low redshifts  in the context of  the $\CC$CDM as compared to observations, the discrepancy  being here  at the more modest  level of $\sim 2-3\sigma$.  A disruption in the opposite direction  has recently appeared from the LSS data collected by the James Webb Space Telescope (JWST)\,\cite{Gardner:2006ky,Labbe:2022ahb}, which  has discovered populations of very massive galaxies at large redshifts $z\gtrsim 10$, a fact which is completely unexpected within the $\CC$CDM model.

The myriad strategies concocted in the literature  to cope with the above phenomenological tensions have nevertheless not been able to fix the situation completely, and  since the tensions do not seem to fade away in the scientific  horizon after years of struggle,  this topic  remains as one of the main focuses of current cosmological research.  We refer once more the reader to the above mentioned reviews for comprehensive lists of references dealing with multifarious models.  In our opinion, however,  the problem with the tensions and the CCP are not necessarily independent issues, and as shown in specific analyses\,\cite{SolaPeracaula:2021gxi,SolaPeracaula:2023swx}, frameworks  capable of alleviating the CCP (such as the RVM) have a bearing on fixing the tensions as well.  Herein we shall exclusively focus on the possibility that the DE is a composite DE fluid, which represents a real novelty to tackle the tensions, despite some composite models existed already in the past aiming at different purposes (such as the cosmic coincidence problem).  A well-known example is the composite running vacuum model  $\CC$XCDM studied in \cite{Grande:2006nn,Grande:2006qi,Grande:2008re}.  A recent work inspired along these lines \cite{Gomez-Valent:2024tdb} has put forward the idea of the $w$XCDM model (not to be confused with the standard $w$CDM parameterization\cite{Turner:1997npq}). The $w$XCDM is a nice prototype of ` transitional composite model' which proves extremely effective in cutting down the tensions under specific conditions, especially when data on angular BAO are employed in the fitting analysis \cite{Gomez-Valent:2023uof}.  In it, we have two components $(X,Y)$ which  are subject to dynamical evolution, each one having a different equation of state (EoS), denoted  $(w_X,w_Y)$. The $Y$ component behaves effectively as a dynamical vacuum term and hence it mimics the RVM, since in the latter the  VED is dynamical and, in addition, it possesses a dynamical EoS slightly departing from $-1$\,\cite{Moreno-Pulido:2022upl}.  Component $X$, on the other hand,  can play the exotic role of ``phantom matter'' (PM) \cite{Grande:2006nn}, namely an intriguing form of DE which, in contradistinction  to the usual phantom DE, is characterized by positive pressure ($p_X>0$) at the expense of a negative energy density ($\rX<0$).  An example of fundamental PM behavior appears in the stringy RVM context of \cite{Mavromatos:2020kzj,Mavromatos:2021urx}.  A model which can be viewed as a particular case of the $w$XCDM appears when  $w_X$ and $w_Y$ are both stuck to $-1$ and one assumes a transition from AdS to dS,  a situation which was studied in \cite{Akarsu:2023mfb,Akarsu:2021fol} and called the  $\CC_s$CDM model. It has no flexibility to accommodate dynamical DE though. It is important to emphasize that components $X$ and $Y$ must act chronologically in sequence during the late cosmic evolution (i.e. after decoupling of the CMB):  namely, $X$  acts first  until a transition redshift $z_t$  near our time, and  below that redshift $Y$ takes over. The latter is the only DE component that shows its face to us since  it is the one that reaches up to our days and is responsible for the cosmic acceleration that we observe.

Upon fitting the three parameters  $(w_X,w_Y,z_t)$ of the $w$XCDM to a large set of cosmological data we find the following:  i) the transition point is relatively close: $z_t\sim 1-2$,  ii) component $Y$ behaves  as quintessence ($w_Y\gtrsim-1$), a result which is compatible with the recent DESI results\cite{DESI:2024mwx,DESI:2024aqx,DESI:2024hhd}; and iii) the component $X$ behaves as PM (this means $w_X\lesssim-1$ but also, as indicated, $\rho_X<0$).  Now since $X$ has a negative energy density,  the observed  value of $H_0$ in the low redshift range, which is governed by the $Y$ component,  must be  larger (closer to the SH0ES value \cite{Riess:2021jrx})  in order to compensate for the fact that  the PM component $X$  makes the distance from $z_t$ to the last scattering surface larger.  Thus, as a result of the composite $(X,Y)$ structure  of the  $w$XCDM model,  the $H_0$ tension can be fixed from the interplay between the PM and quintessence behaviors of the two components of the DE.  At the same time the growth tension may be highly alleviated since in the recent universe (ruled by  the $Y$ component of the DE) the structure formation is slowed down under quintessence-like behavior, as it will be shown in our study.  But by the same token the $X$ component of the DE, which carries positive pressure $p_X=w_X\rho_X>0$ and  is in force  at the upper redshift range $z>z_t$, may on its own produce unsuspected large scale structures at high redshifts, providing a chance to explain the aforementioned JWST observations\,\cite{Labbe:2022ahb}.

In this Letter, we further explore the $w$XCDM and shall present the main fitting results with a  more recent set of SNIa as compared to \cite{Gomez-Valent:2024tdb}. In addition, we analyze for the first time the differences between using  BAO 2D versus  BAO 3D data along with the remaining set of observations, which involve cosmic chronometers, LSS data, the SNIa calibration from SH0ES and the full CMB likelihood from Planck 2018. Finally, we compare the behavior of the $w$XCDM, not only  with the standard  $\CC$CDM model, but also with the mentioned $\CC_s$CDM and with our own fitting results obtained with the  {$w$CDM and $w_0w_a$CDM parameterizations} of the DE used recently by DESI\cite{DESI:2024mwx,DESI:2024hhd}. The $w_0w_a$CDM is sometimes also called the CPL parameterization\,\cite{Chevallier:2000qy,Linder:2002et}. We find that  $w$XCDM  and $w_0w_a$CDM provide a very efficient description of the observations, both with BAO 2D and BAO 3D, but it is only the composite model $w$XCDM that provides the optimal fitting results with BAO 2D; and in this context the tensions can be fully accounted for.


\section{Monocomponent versus composite dark energy}\label{sec:CompositeDE}
`Ordinary' formulations of the DE are monocomponent, i.e. they involve just one cosmic substance or fluid. Thus, in the standard  $\CC$CDM the single  `fluid'  is the VED associated with the CC term $\CC$.  Similarly, DE proposals such as quintessence or phantom DE\,\cite{Peebles:2002gy,Padmanabhan:2002ji,Copeland:2006wr} are usually monocomponent since they involve a single scalar field.  However, the DE can also be multicomponent or composite, i.e. with two or more cosmic components or fluids.  For example, in some formulations (take  the old quintom model\,\cite{Feng:2004ad}, for instance)  quintessence and phantom DE components may participate together. Herein we do not wish to consider scalar field models of the DE, as they are committed to specific forms of the effective potential (s).  In what follows, we instead focus on generic forms of the DE of unspecified nature, both monocomponent and composite,  which in some cases are grounded on fundamental physics (such as QFT or string theory).  In doing this we wish to compare  usual parameterizations of the DE, such as  $w$CDM\,\cite{Turner:1997npq} or $w_0w_a$CDM\,\cite{Chevallier:2000qy,Linder:2002et} (usually presuming  monocomponent scenarios) with alternative parameterizations mimicking composite scenarios of the DE, specifically those involving the possibility of a sign flip of the DE density. As we shall see, this sign flip feature can be particularly efficient to improve the overall fit quality of the cosmological observations and can be instrumental for alleviating the cosmological tensions\cite{Akarsu:2023mfb,Akarsu:2021fol,Gomez-Valent:2023uof,Gomez-Valent:2024tdb,SolaPeracaula:2024iil}.  In the following, we mention a few models of the DE, one of them being a pretty standard parameterization of the dynamical DE usually based on assuming an underlying  monocomponent fluid, and the other three incorporate the sign flip feature:
\vspace{0.5cm}

$\bullet$ i)  $w_0w_a$CDM (or CPL) parameterization \cite{Chevallier:2000qy,Linder:2002et} of the DE, characterized by the following dynamical EoS:
\begin{equation}\label{eq:CPL}
 w(z)=w_0+w_a(1-a)=w_0+\frac{w_a z}{1+z}\,,
\end{equation}
which is given either in terms of the scale factor or the redshift ($1+z=1/a$).
A particular case is the $w$CDM parameterization\,\cite{Turner:1997npq}, in which $w_0$ is simply called $w$ and  $w_a=0$, hence the EoS in this situation is non-dynamical (although, of course, the DE is still evolving with the expansion). In our analysis, we will also provide separate results corresponding to this simpler parameterization, together with the more general form \eqref{eq:CPL}  used by the DESI Collaboration \cite{DESI:2024mwx,DESI:2024hhd}.
The corresponding Hubble rate $H=\dot{a}/a$ for the more  genereal EoS  \eqref{eq:CPL} can be readily derived:
\begin{equation}
H(a)=H_0\left[(\Omega^0_b+\Omega^0_{cdm}){a^{-3}} + \Omega_\gamma^0{a^{-4}}+ \frac{\rho_\nu(a)}{\rho_{c 0}} + \Omega_{\rm DE}^0\,{a^{-3(1+w_0+w_a)}}e^{-3w_a(1-a)}\right]^{1/2}\,.
\end{equation}
Here $H_0$ is the current value of the Hubble rate, and $\Omega^0_i=\rho_i^0/\rho_c^0$ are today's energy densities of baryons, cold dark matter, photons and  DE normalized with respect to the current critical density. We note that the exact neutrino contribution,  $\rho_\nu(a)$, cannot be expressed analytically since it contains a massive component, $\rho_{\nu, m}(a)$, apart from the massless relativistic one with the usual behavior $\rho_{\nu,r}\sim a^{-4}$. Therefore, during the expansion of the universe, the neutrino transits from a relativistic into a nonrelativistic regime.  Being this process nontrivial, it has to be dealt with numerically. In this work, we use in all cases one massive neutrino of $0.06$ eV and two massless neutrinos.

$\bullet$ ii) $\CC$XCDM model\cite{Grande:2006nn,Grande:2006qi,Grande:2008re}. In this case, the cosmic fluid contains, in addition to the usual matter energy density $\rmr$,  a composite DE sector constituted of two subcomponents, to wit: one is the running vacuum energy density $\rv$, and the other is an effective entity called $X$, whose energy density is $\rX$. Since the model is formulated within the RVM framework\cite{SolaPeracaula:2022hpd}, the evolution of the VED can be determined within the QFT formalism of \cite{Moreno-Pulido:2020anb,Moreno-Pulido:2022phq,Moreno-Pulido:2022upl,Moreno-Pulido:2023ryo}, in which $\rv= \rv(H)$ evolves with the expansion rate.  In the present universe, the VED evolution adopts the RVM form
\begin{equation}\label{eq:RVM}
\rv(H) = \rvo+\frac{3\nueff}{8\pi}\,(H^2-H_0^2)\,\mpl^2\,,
\end{equation}
with $\rvo$  the current VED value,  $\mpl$  the Planck mass  and  $|\nueff|\ll1$  a small  parameter which can be computed in QFT, see the aforementioned papers. For $\nueff>0$ the VED decreases with expansion and the model mimics quintessence, while for $\nueff<0$ the VED increases with the expansion and in this alternative situation the model behaves effectively as phantom DE. In the original version of $\CC$XCDM\cite{Grande:2006nn}, there is an exchange of energy between the VED and the $X$ component, and as a result the solution of the cosmological equations in terms of the scale factor or the cosmological redshift is a bit cumbersome despite being fully analytic. The study of cosmic tensions within the $\CC$XCDM framework will be presented elsewhere. A simplified version of this model, which nevertheless emulates the basic features of the latter, is the $w$XCDM model defined in the next point.

$\bullet$  iii) \wXCDM model\cite{Gomez-Valent:2024tdb}.  In this model, we have two dynamical components for the DE, denoted $(X,Y)$, with respective EoS parameters ($w_X,w_Y)$.  The component $X$ has similarities to the aforementioned $\CC$XCDM, whereas $Y$ mimics the dynamical behavior of $\CC$ in the $\CC$XCDM.  Upon fitting the model to the data, the $X$ component is phantom-like ($\wX\lesssim -1$)  while the $Y$ component is quintessence-like ($\wY\gtrsim -1$). However, $X$ does not behave as usual phantom DE since its energy density is negative ($\rho_X<0$) and therefore its pressure is positive, $p_X=w_X\rho_X>0$, which is why it is called `phantom matter' (PM) -- first introduced in the context of the aforementioned $\CC$XCDM model\cite{Grande:2006nn}\footnote{We refer the reader to Sec. 3 and Fig. 1 of \cite{Gomez-Valent:2024tdb} for a clear discussion and graphical illustration of the position of phantom matter in the context of the energy conditions satisfied by the various cosmic fluids. }. Despite we will treat here PM in a pure phenomenological manner, let us recall that there are fundamental scenarios where the existence of PM can be substantiated, e.g. in stringy formulations of the RVM\,\cite{Mavromatos:2020kzj,Mavromatos:2021urx}. As mentioned in the Introduction, it is important to clarify that $X$ and $Y$ do not act simultaneously along the cosmic evolution: $X$ acts first until a transition redshift $z_{t}$ (fitted from the data), i.e. for $z>z_t$, while $Y$ acts below that redshift ($z<z_t$) until the present day (see Fig. \ref{fig:background}). The form of the energy densities and pressures of the $Y$ and $X$ components read, respectively,

\begin{equation}
\rho_Y(z) = \rho_Y^0(1+z)^{3(1+w_Y)}\qquad ;\qquad p_Y(z)=w_Y\rho_Y(z)\,,
\end{equation}
and

\begin{equation}
\rho_X(z) = -\rho_Y^0(1+z_t)^{3(w_Y-w_X)}(1+z)^{3(1+w_X)}\qquad ;\qquad p_X(z)=w_X\rho_X(z)\,,
\end{equation}
with $\rho_Y^0$ the current value of the energy density of $Y$.  Notice that the instantaneous jump at $z_t$ is only a $\theta$-function approximation of a more complex and continuous process that brings the universe from the PM phase to the quintessence phase. A detailed description would necessarily introduce more parameters and hence  would be model-dependent. The jump must be considered as a phenomenological simplification that keeps the number of free parameters to a minimum while still encapsulating the most important features of the model.

In the above description, the $Y$ component  is the `visible' face of the composite DE fluid. In fact, the $Y$ component has positive density $\rho_Y>0$ and negative pressure $p_Y<0$. Since its EoS is quintessence-like ($\wY\gtrsim -1$), as noted, this component appears to us as a pretty `standard' DE fluid. What about the hidden face of the composite DE, i.e. the PM component $X$? Although it only reigns over far domains of our past, specifically for $z>z_t\sim 1.4$ (see Sec.\,\ref{sec:NumericalAnalysis} for details), and hence is not directly accessible to us,  it nevertheless has dramatic consequences for improving the overall fit quality to the cosmological data, as we shall see in our analysis. We should immediately clarify at this point that the free parameters of the $w$XCDM model are just three: $(z_{t}, w_X, w_Y)$. The density parameters for the DE components are not independent degrees of freedom, since, e.g. the value of $\rho_Y^0\equiv\rho_Y(z=0)$ can be obtained from the fitting values of the standard parameters $H_0, \omega_b, \omega_{\rm dm}$.
Furthermore, it is assumed that the condition $|\rho_X(z)|=\rho_Y(z)$ holds at the transition redshift  $z=z_{t}$, so the values of the two DE densities $\rho_X$ and $\rho_Y$  on both sides of $z_{t}$ (i.e. just above and below this value) are taken to be equal in absolute value but opposite in sign, which induces a discontinuity in the Hubble function, but keeps the cosmological distances continuous.  This assumption is intended to reduce the number of free parameters in composite models with a sign flip of the DE density.

$\bullet$ iv) $\CCS$ model\cite{Akarsu:2023mfb}.  As indicated in the introduction, this model can be conceived as a particular case of the $w$XCDM, in which  $X$ and $Y$ are assumed to be rigid (non-evolving) $\CC$-like components: $w_X=w_Y=-1$. The model involves a transition redshift $z_t$,  above which the $X$ component forms an anti-de Sitter (AdS) phase ($\CC<0$) and below it the $Y$ component  behaves as a standard (positive) cosmological constant (i.e. a de Sitter phase, $\CC>0$). It is also assumed that there is a sign flip of $\CC$ at $z=z_t$ which preserves the absolute value of the DE  (cf. Fig. \ref{fig:background}). Strictly speaking, $\CCS$ is not a genuine composite model, although we will still refer to it as such.

In the following, we will furnish detailed numerical analyses of  five models of the DE, to wit: $w$CDM and $w_0w_a$CDM mentioned in i), the basic composite model $w$XCDM described in iii), and  the particular case $\CCS$  indicated in iv).  At the same time, we will provide the corresponding results for the standard $\CC$CDM.  All five models will be confronted to the same set of cosmological data and compared statistically to the $\CC$CDM. Thus, in the present study we shall assess the impact of composite scenarios iii) and iv) against monocomponent scenarios, represented here by the two conventional parameterizations in i) and the $\CC$CDM model.

\section{Data}\label{sec:Data}

We constrain the $w$XCDM model as well as the benchmark models described in Sec. \ref{sec:CompositeDE} using a very similar methodology to the one employed in our previous work \cite{Gomez-Valent:2024tdb}. In particular, we use: the full Planck 2018 CMB temperature, polarization and lensing likelihoods \cite{Planck:2018vyg}; 33 data points on $H(z)$  from cosmic chronometers (CCH) in the redshift range $z<2$ \cite{Jimenez:2003iv,Simon:2004tf,Stern:2009ep,Zhang:2012mp,Moresco:2012jh,Moresco:2015cya,Moresco:2016mzx,Ratsimbazafy:2017vga,Borghi:2021rft,Tomasetti:2023kek}; and galaxy clustering data extracted from the analysis of redshift-space distortions (RSD) and peculiar velocities \cite{Guzzo:2008ac,Song:2008qt,Blake:2011rj,Blake:2013nif,Simpson:2015yfa,Gil-Marin:2016wya,eBOSS:2020gbb,Said:2020epb,Avila:2021dqv,Mohammad:2018mdy,Okumura:2015lvp}. As in \cite{Gomez-Valent:2024tdb}, we take into account the CCH non-diagonal covariance matrix \cite{Moresco:2020fbm} and treat the LSS data in terms of the observable $f(z)\sigma_{12}(z)$, which is more robust than the usual $f(z)\sigma_{8}(z)$ since the former encapsulates the clustering information at a fixed scale of $R_{12}=12$ Mpc, which is independent from $h$ and, therefore, less prone to biases \cite{Sanchez:2020vvb,eBOSS:2021poy,Semenaite:2022unt,Forconi:2025}. These subsets of data coincide exactly with those employed in \cite{Gomez-Valent:2024tdb}, but in this paper we make important updates both in the SNIa and BAO samples, namely: (i) we replace the Pantheon+ SNIa compilation \cite{Brout:2022vxf} with the recent full five-year SNIa dataset from the Dark Energy Survey, referred to as DES-Y5 \cite{DES:2024jxu,DES:2024hip}. This sample has proved to be more sensitive to dark energy dynamics, see, e.g. \cite{DESI:2024mwx,DESI:2024hhd}. We also use a prior on the absolute magnitude of SNIa, as measured by the SH0ES Team in the first two rungs of the cosmic distance ladder, $M=-19.253\pm  0.027$ mag \cite{Riess:2021jrx}; and (ii) we combine the CMB+CCH+LSS+DESY5+SH0ES data set with two different types of BAO measurements separately, the so-called transverse (aka angular or 2D) BAO data from Refs. \cite{Carvalho:2015ica,Alcaniz:2016ryy,Carvalho:2017tuu,deCarvalho:2017xye,deCarvalho:2021azj} and anisotropic (or 3D) BAO data from Refs. \cite{Carter:2018vce,Kazin:2014qga,Gil-Marin:2016wya,eBOSS:2020uxp,duMasdesBourboux:2020pck}. They complete the two data sets used in this work, to wit: CMB+CCH+LSS+DESY5+SH0ES+BAO\_2D and CMB+CCH+LSS+DESY5+SH0ES+BAO\_3D. The 2D BAO measurements are extracted from the raw tracer maps in the space of angles and redshifts without requiring the use of a fiducial model to convert redshifts into distances. They are, in principle, less affected by model dependencies, which can have a significant impact on 3D analyses \cite{Anselmi:2018vjz,Anselmi:2022exn}. Despite being extracted from the same or very similar catalogs of tracers, these two types of BAO data are obtained following different methodologies and turn out to be in tension \cite{Favale:2024sdq}. Indeed, they lead to completely different solutions to the Hubble tension \cite{Camarena:2019rmj,Gomez-Valent:2023uof}, see also \cite{Bernui:2023byc,Akarsu:2023mfb,Anchordoqui:2024gfa,Gomez-Valent:2024tdb,Dwivedi:2024}. It is, therefore, of utmost importance to analyze the models with 2D and 3D BAO separately in order to understand their impact on the discussion of the cosmological tensions. This is one of the main focuses of the current study.


\section{Numerical analysis and discussion}\label{sec:NumericalAnalysis}

We use a modified version of the Einstein-Boltzmann code \texttt{CLASS} \cite{Lesgourgues:2011re,Blas:2011rf} to solve the coupled system of cosmological background and linear perturbation equations and compute all the theoretical predictions of the models under study. We employ \texttt{MontePython} \cite{Audren:2012wb,Brinckmann:2018cvx} to run the Monte Carlo analyses and test the convergence of the chains with the help of the Gelman-Rubin criterion \cite{GelmanRubin}. We stop the runs when $R-1<0.02$. The Python code \texttt{GetDist} \cite{Lewis:2019xzd} is used to analyze the chains and obtain constraints on the model parameters. Our main fitting results for the various models and for both types of BAO (2D and 3D)  are displayed in the upper and lower parts of Table \ref{tab:table_fits}, respectively. Furthermore, for the convenience of the reader, in Table \ref{tab:table_chi2} of the Appendix A we provide the specific contributions of each observable entering our analysis to the total  $\chi^2_{\rm min}$ for the two fits, one using  BAO 2D-only  and the other  BAO 3D-only. For lack of space, in this Letter we omit the full triangle contour plots for the various models studied in this paper under the two BAO sorts, which display the two-dimensional marginalized likelihood distributions for each parameter. They are similar to those already presented in \cite{Gomez-Valent:2024tdb}.

As is well known, to compare the fitting performance of the various models more fairly, it is necessary to penalize the use of additional parameters, which is tantamount to implementing Occam's razor. This is why in Table \ref{tab:table_fits} we not only report the minimum values of $\chi^2$ obtained for each model, but we also display the differences between the deviance (DIC) \cite{DIC} and Akaike (AIC) \cite{Akaike} information criteria found between the $\Lambda$CDM and the various mono- and multicomponent DE models, expressed in our case as $\Delta{\rm DIC}\equiv{\rm DIC}_{\Lambda{\rm CDM}}-{\rm DIC}_{i}$ and $\Delta{\rm AIC}\equiv{\rm AIC}_{\Lambda{\rm CDM}}-{\rm AIC}_{i}$, respectively, with $i$ referring to any of the models $w$XCDM, $\Lambda_s$CDM, $w$CDM or $w_0w_a$CDM. The DIC itself is defined as

\begin{equation}\label{eq:DIC}
   {\rm DIC} =  \chi^2(\bar{\theta})+2p_D\,,
\end{equation}
with $p_D=\overline{\chi^2}-\chi^2(\bar{\theta})$ the effective number of parameters in the model, $\overline{\chi^2}$ the mean value of $\chi^2$ and $\bar{\theta}$ the mean of the parameters entering the Monte Carlo analysis. It incorporates the information encapsulated in the full Markov chains. Similarly, when the number of points is much larger than the number of parameters $n_p$ (which is certainly  the case here), the AIC is defined as

\begin{equation}\label{eq:AIC}
    {\rm AIC} =  \chi^2_{\rm min}+2n_p\,.
\end{equation}
With the above definitions, a positive difference in these information criteria indicates a better performance of the composite/dynamical DE models than the standard $\Lambda$CDM. If $0 \leq \Delta\textrm{DIC}<2$ it is said that one finds \textit{weak evidence} in favor of the new model under test, compared to the standard model. If $2 \leq \Delta\textrm{DIC} < 6$, we speak instead of \textit{positive evidence}. If $6 \leq \Delta\textrm{DIC} < 10$, there is \textit{strong evidence} in favor of the composite DE models, whilst if  $\Delta\textrm{DIC}>10$ we can conclude that there is \textit{very strong evidence} supporting the new model against the standard $\CC$CDM. An analogous consideration can be made using AIC, of course. Despite the simplicity of AIC as compared to DIC, we obtain respective values for these two information statistics that are fully consistent.

\begin{table}[t!]
\centering
\renewcommand{\arraystretch}{1.76}
\resizebox{\textwidth}{!}{
\begin{tabular}{|c ||c |c | c | c  |c |}
\hline
\multicolumn{6}{|c|}{\textbf{CMB+CCH+SNIa+SH0ES+BAO\_2D+$f\sigma_{12}$}} \\
\hline
\hline
\hline
{\small Parameter} & {\small $\Lambda$CDM} & {\small $w$CDM } & {\small $w_0w_a$CDM}    & {\small $w$XCDM }  & {\small $\Lambda_s$CDM}
\\\hline
$10^2\omega_b$ & $2.271\pm 0.014$ (2.267) & $2.269\pm 0.014 $ (2.265) & $2.257^{+0.012}_{-0.014}$ (2.251) & $2.249\pm 0.012$ (2.249) & $2.236\pm 0.017$ (2.221)  \\\hline
$10\,\omega_{\rm dm}$ &  $1.160\pm 0.009$ (1.159)& $1.162\pm 0.010$ (1.175) & $1.177\pm 0.010$ (1.177) & $1.185\pm 0.009$ (1.185)  & $1.203\pm 0.016$ (1.223)  \\\hline
$\ln(10^{10}A_s)$ & $3.060^{+0.016}_{-0.018}$ (3.067) & $3.059^{+0.014}_{-0.017}$ (3.042) & $3.041\pm 0.013$ (3.050) & $3.042^{+0.014}_{-0.016}$ (3.037)  &  $3.036\pm 0.015$ (3.025)  \\\hline
$\tau$ & $0.066^{+0.008}_{-0.009}$ (0.068) & $0.065^{+0.007}_{-0.009}$ (0.056) & $0.055\pm 0.007$ (0.061) & $0.054\pm 0.008$ (0.053)  &  $0.050\pm 0.008$ (0.044)   \\\hline
$n_{s}$ &  $0.975\pm 0.004$ (0.974) & $0.974\pm 0.004$ (0.972) & $0.970\pm 0.004$ (0.975) & $0.969\pm 0.004$ (0.968)   &  $0.965\pm 0.005$ (0.960)   \\\hline
$H_{0}$  &  $69.28\pm 0.41$ (69.22) & $69.40\pm 0.63$ (69.37) & $69.77\pm 0.58$ (69.56) & $70.94\pm 0.56 $ (71.29)   & $72.36\pm 0.91$ (73.18)  \\\hline
$z_t$ & $-$ & $-$ & $-$ & $1.46\pm 0.02$ (1.44)  & $1.70^{+0.16}_{-0.27}$ (1.46) \\\hline
$w_X$ & $-$ & $-$ & $-$ & $-1.67\pm 0.23$ (-1.80)  &  $-$ \\\hline
$w_Y$ & $-$ & $-$ & $-$ & $-0.859\pm 0.028$ (-0.882) & $-$ \\\hline
$w_0$ & $-$ & $-1.008^{+0.019}_{-0.022}$ (-1.024) & $-0.550^{+0.084}_{-0.043}$ (-0.528) & $-$  &  $-$ \\\hline
$w_a$ & $-$ & $-$ & $-2.04^{+0.16}_{-0.39}$ (-2.10) & $-$ & $-$ \\\hline\hline
$\Omega_m^0$ & $0.290\pm 0.005$ (0.291) & $0.290\pm 0.006$ (0.293) & $0.290\pm 0.006$ (0.291) & $0.282\pm 0.005$ (0.279) & $0.274\pm 0.006$ (0.271) \\\hline
$M$ &  $-19.360\pm 0.010$ (-19.362) & $-19.359\pm 0.013$ (-19.361) &  $-19.285^{+0.016}_{-0.013}$ (-19.289) & $-19.274\pm  0.013$ (-19.271) & $-19.279\pm 0.024$ (-19.253) \\\hline
$\sigma_{12}$ & $0.784\pm 0.007$ (0.787) & $0.786\pm 0.008$ (0.788) & $0.798\pm 0.008$ (0.803) &   $0.776\pm 0.007$ (0.774) & $0.785\pm 0.007$ (0.789)
 \\\hline\hline
$\chi^2_{\rm min}$ &  $4523.64$ & $4523.60$ & $4472.42$ & $4457.84$ &  $4488.56$ \\\hline
$\Delta$DIC &  $-$ & -0.42 & 49.31 & 63.19  & 32.28 \\\hline
$\Delta$AIC &  $-$ & -1.96 & 47.22 & 59.80 & 33.08 \\\hline\hline\hline
\multicolumn{6}{|c|}{\textbf{CMB+CCH+SNIa+SH0ES+BAO\_3D+$f\sigma_{12}$}} \\
\hline\hline
\hline
{\small Parameter} & {\small $\Lambda$CDM} & {\small $w$CDM } & {\small $w_0w_a$CDM}    & {\small $w$XCDM }  & {\small $\Lambda_s$CDM}\\\hline
$10^2\omega_b$ & $2.262\pm 0.012$ (2.255) & $2.267\pm 0.013$ (2.259) & $2.251\pm 0.014$ (2.248) & $2.244^{+0.012}_{-0.014}$ (2.246) &  $2.239\pm 0.013$ (2.239) \\\hline
$10\,\omega_{\rm dm}$ & $1.175\pm 0.007$ (1.181) & $1.170\pm 0.009$ (1.169) & $1.188\pm 0.009$ (1.190) & $1.198\pm 0.011$ (1.189) & $1.203\pm 0.010$ (1.200) \\\hline
$\ln(10^{10}A_s)$ & $3.049^{+0.013}_{-0.016}$ (3.051) & $3.057\pm 0.016$ (3.065) & $3.037\pm 0.014$ (3.036) & $3.039\pm 0.013$ (3.036)  &  $3.034\pm 0.014$ (3.038) \\\hline
$\tau$ & $0.059^{+0.006}_{-0.008}$ (0.058) & $0.064\pm 0.008$ (0.067) & $0.052\pm 0.007$ (0.051) & $0.052\pm 0.007$ (0.053) &  $0.049\pm 0.007$ (0.049)  \\\hline
$n_{s}$ & $0.971\pm 0.004$ (0.969) & $0.973\pm 0.003$ (0.976) & $0.968\pm 0.003$ (0.964) & $0.966\pm  0.004$ (0.968) &  $0.964\pm 0.004$ (0.967)  \\\hline
$H_{0}$ & $68.58\pm 0.32$ (68.33) & $68.07\pm 0.56$ (68.09) & $67.61\pm 0.51$ (67.12) & $68.43\pm 0.52$ (68.04) & $69.39\pm 0.37$ (69.69) \\\hline
$z_t$ & $-$ & $-$ & $-$ & $2.08^{+0.26}_{-0.30}$ (2.00) & $2.32^{+0.15}_{-0.28}$ (2.22) \\\hline
$w_X$ & $-$ & $-$ & $-$ & $<-1.08$ (-1.59)  &  $-$ \\\hline
$w_Y$ & $-$ & $-$ & $-$ & $-0.933\pm 0.025$ (-0.906) & $-$ \\\hline
$w_0$ & $-$ & $-0.976\pm 0.022$ (-0.979) & $-0.762\pm 0.039$ (-0.735) & $-$  &  $-$ \\\hline
$w_a$ & $-$ & $-$ & $-0.87^{+0.14}_{-0.16}$ (-0.93) & $-$ & $-$ \\\hline\hline
$\Omega_m^0$ & $0.299\pm 0.004$ (0.303) & $0.303\pm 0.005$ (0.302) & $0.311\pm 0.005$ (0.315) & $0.305\pm 0.005$ (0.307) & $0.298\pm 0.004$ (0.294) \\\hline
$M$ & $-19.376\pm 0.009$ (-19.383) &  $-19.383\pm 0.011$ (-19.381) & $-19.362\pm 0.011$ (-19.367) & $-19.358\pm 0.011$ (-19.365) & $-19.352\pm 0.010$ (-19.345) \\\hline
$\sigma_{12}$ & $0.790\pm 0.006$ (0.794) & $0.788\pm 0.007$ (0.793) & $0.797^{+0.007}_{-0.006}$ (0.798) & $0.791\pm 0.007$ (0.786) & $0.793\pm 0.006$ (0.793)
 \\\hline\hline
$\chi^2_{\rm min}$ & 4514.84 & 4514.20 & 4493.92 & 4497.46 &  4504.92 \\\hline
$\Delta$DIC &  $-$ & -1.26 & 16.94 & 15.36 & 10.15 \\\hline
$\Delta$AIC &  $-$ & -1.36 & 16.92 & 11.38 & 7.08 \\\hline
\end{tabular}}
\caption{\scriptsize Mean values and uncertainties at 68\% CL obtained with the full data sets CMB+CCH+SNIa+SH0ES+BAO\_2D+$f\sigma_{12}$ (upper half) and CMB+CCH+SNIa+SH0ES+BAO\_3D+$f\sigma_{12}$ (lower half). We show the best-fit values in brackets. We use the standard notations for the $\CC$CDM parameters, and $H_0$ is given in km/s/Mpc. In the last three lines of the two subtables, we display the values of the minimum $\chi^2$, $\Delta$DIC and $\Delta$AIC, as defined in Eqs. \eqref{eq:DIC} and \eqref{eq:AIC}, respectively.}
\label{tab:table_fits}
\end{table}

\begin{figure}[t!]
    \centering
    \includegraphics[scale=0.75]{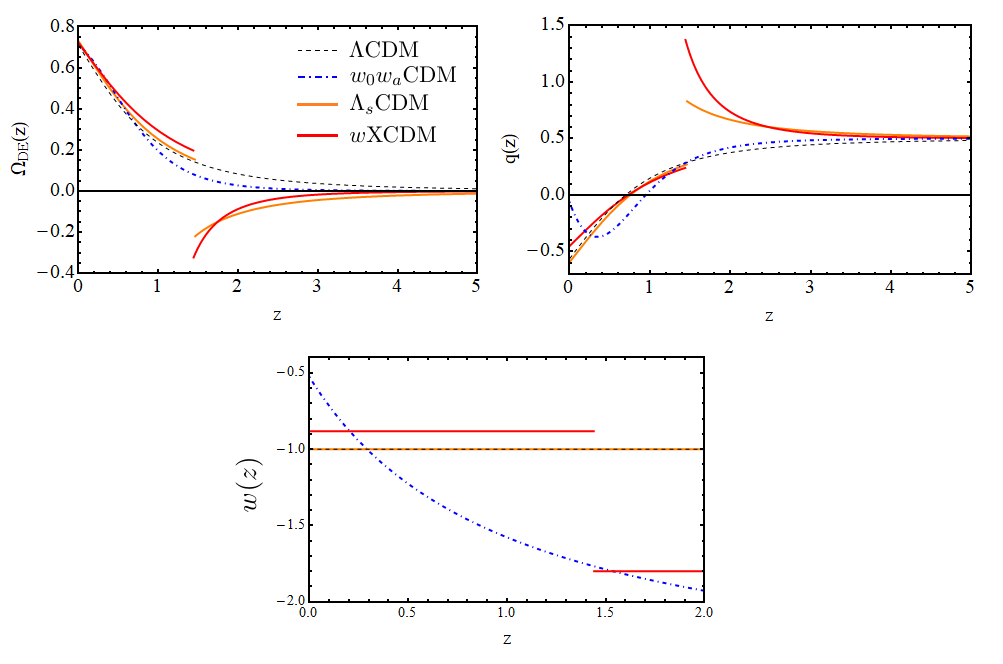} \caption{\scriptsize {\it Left upper plot:} Dark energy fraction $\Omega_{\rm DE}(z)=\frac{\rho_{\rm DE}(z)}{\rho_c(z)}$ as a function of the redshift for $\Lambda$CDM, $w_0w_a$CDM, $\Lambda_s$CDM and $w$XCDM, obtained using the best-fit parameters with CMB+CCH+LSS+DESY5+SH0ES+BAO\_2D data (cf. Table \ref{tab:table_fits}); {\it Right upper plot:} The corresponding curves for the deceleration parameter, $q=-\ddot{a}a/\dot{a}^2$; {\it Lower plot:} Evolution of the EoS parameter of the various models. For the $w$XCDM, $w_X<-1$ at $z>z_t$ and $X$ behaves as phantom matter. At $z<z_t$, $w_Y>-1$ and $Y$ evolves as quintessence. For $\Lambda$CDM and $\Lambda_s$CDM, $w=-1$, and for $w_0w_a$CDM  we use Eq. \eqref{eq:CPL}.}
    \label{fig:background}
\end{figure}

Let us now first focus on analyzing the results obtained using the data set  with BAO 2D-only. It has been shown that with this BAO type and in models exhibiting a transition from negative to positive dark energy densities around $z_t\sim 1.5-2$, the Hubble tension can be strongly alleviated, if not completely erased\cite{Gomez-Valent:2023uof}.  This has been  explicitly demonstrated in previous works both in the context of the $\Lambda_s$CDM \cite{Akarsu:2023mfb} and the $w$XCDM\cite{Gomez-Valent:2024tdb}. Nevertheless, the overall fitting performance of these two models is by no means the same owing to the different behavior of the DE, namely rigid versus dynamical, respectively. In \cite{Gomez-Valent:2024tdb}, we demonstrated that current data strongly prefers the $w$XCDM model over  $\Lambda_s$CDM, essentially due to a substantial improvement in the description of the SNIa apparent magnitudes from the Pantheon+ compilation, and we already advanced in the conclusions that the analysis with the DES-Y5 sample would probably enhance the differences between these two models in favor of $w$XCDM. Our conjecture gets now fully confirmed in the current study. From the main fitting table, we can see that the values of $\Delta$AIC and $\Delta$DIC are $\sim 30$ units larger in the $w$XCDM than in the $\CC_s$CDM, which points to an outstanding preference for the former model. The combination of BAO 2D and DES-Y5 SNIa with the other data sets (which include the SH0ES prior) forces the value of $\Omega_m^0$ in both models to be pretty small, close to $\sim 0.27-0.28$. This is due to the large (positive) values of the DE density after the transition (viz. $z<z_t$), which are needed to compensate the flip of DE sign occurring at $z=z_t$ (see the left plot of Fig. \ref{fig:background}). This transition is required to properly fit the angle $\theta_*$ (the angular size of the sound horizon) measured by Planck \cite{Planck:2018vyg}, which can be translated into a very precise measurement of the distance to the last scattering surface if the model does not introduce new physics before recombination. This is actually the case in all the models under consideration, for which we find $r_d\sim 147$ Mpc\footnote{Notice that we have not allowed $r_d$ to vary freely in the Monte Carlo analysis, but treat it as a derived parameter. We compute $r_d$ with \texttt{CLASS} in a self-consistent way within the various models.}. As it is obvious from Fig. 7 of \cite{DES:2024jxu}, such low values of the matter fraction are disfavored at $\sim 3\sigma$ CL by DES-Y5 SNIa if DE is non-dynamical, i.e. if $w=-1$ (as in $\Lambda_s$CDM)\footnote{Similar conclusions are drawn from Fig. 18 of the Union 3 paper \cite{Rubin:2023ovl} and, with lower statistical significance, from Fig. 9 of the Pantheon+ paper \cite{Brout:2022vxf}.}. This causes a sizable enhancement of $\chi^2_{\rm SNIa}$ in the $\Lambda_s$CDM fit as compared to the more flexible $w$XCDM fit, see Table \ref{tab:table_chi2}. Indeed, such low values of $\Omega_m^0$ can only accommodate the SNIa data if and only if DE is of quintessence type, with $w\gtrsim -0.9$. This is why $w$XCDM is able to do an excellent job: we obtain $w_Y=-0.86\pm 0.03$ (cf. Table \ref{tab:table_fits}), which lies in the correct region of the parameter space and is fully consistent with the DESI results with a comparable dataset \cite{DESI:2024mwx,DESI:2024hhd}.  $\Lambda_s$CDM offers also a much poorer fit to Planck's CMB data, and this is the second fact that makes this model to perform much worse than $w$XCDM. In both models we obtain fitting values for the transition redshift in the range $z_t\sim 1.5-1.9$, similar to the those reported in \cite{Akarsu:2023mfb,Gomez-Valent:2023uof,Gomez-Valent:2024tdb}.

Regarding the Hubble tension,  within the $w$XCDM we obtain a large value of $H_0$ ($H_0=70.94$ km/s/Mpc), slightly lower though than that obtained within the $\Lambda_s$CDM ($H_0=72.36\pm 0.91$ km/s/Mpc). However, it is important to notice that the SH0ES calibration of the absolute magnitude parameter, $M=-19.253\pm 0.027$, when combined with the apparent magnitudes of the DES-Y5 SNIa in the Hubble flow, leads to a measurement of $H_0=70.5\pm 1.1$ km/s/Mpc \cite{Favale:2025}, which is indeed fully compatible with the posterior value that we have obtained for the $w$XCDM (within only $0.36\sigma$),  while in contrast lies $\sim 1.3\sigma$ below the $\Lambda_s$CDM value\footnote{In this work we compare the posterior values of $H_0$ obtained in the various models and fitting analyses with the determination obtained by combining the SH0ES calibration of $M$ and the SNIa in the Hubble flow from DES-Y5, which are the ones that we are using in this study. This is more consistent than making the comparison using the Pantheon+SH0ES measurement, $H_0=73.04\pm 1.04$ km/s/Mpc \cite{Riess:2021jrx}. The shift in the value of $H_0$ extracted from the distance ladder could be caused by systematics either in the low-z SNIa of Pantheon+ or DES-Y5 \cite{Efstathiou:2024xcq,Notari:2024zmi,DES:2025tir}.}. Moreover, no  tension is observed between the SH0ES calibration of $M$ and the values obtained for this nuisance parameter in the two models. Using the $w_0w_a$CDM parameterization, there is no sizeable tension either, whereas within $\Lambda$CDM the tension in $M$ reaches $\sim3.7\sigma$ CL. The take-home message that ensues from our analysis is that the Hubble tension is basically washed out in the transitional composite DE models and the $w_0w_a$CDM when BAO 2D is employed in the fitting analyses. Now, in terms of global fit quality, the balance is much better for the composite $w$XCDM model. Indeed, among the dynamical models fitted using BAO 2D-only data,  we find that the information criteria clearly select $w$XCDM in the first place, followed by $w_0w_a$ CDM at considerable `distance' in AIC and DIC space, and finally $\CC_s$CDM in the third position (see the upper half of Table \ref{tab:table_fits}). The three of them do fit however  the data substantially better than the $\CC$CDM. This positive aspect of the main models is in  stark contrast to the simplest $w$CDM parameterization, where the EoS is not dynamical and this does not allow to have any  improvement whatsoever with respect to the $\CC$CDM, as it is manifest in Table \ref{tab:table_fits}.

At the end of the day, it turns out that the standard $\Lambda$CDM becomes extremely disfavored compared to all other dynamical models (except $w$CDM) in light of the large CMB+CCH+LSS +DESY5+ SH0ES+BAO\_2D dataset. It is true that BAO 2D also allows $H_0$ to increase within the standard $\Lambda$CDM, reaching in this case $H_0\sim 69.3$ km/s/Mpc. However, this boost is not as efficient as in the transitional DE models because, as mentioned above,  the value of $M$ remains in $3.7\sigma$ tension with SH0ES. In addition, $\CC$CDM has to pay a high price to make this increase possible; for, as is apparent from Table \ref{tab:table_chi2} in Appendix A, the $\Lambda$CDM worsens the description of all the individual data sets compared to the other models.  In particular, we can see in that table that the increase of $H_0$ is accompanied by a significant enhancement of $\chi^2_{\rm CMB}$, of about 10 units compared to the value obtained when BAO 3D is used instead of BAO 2D.  As discussed in \cite{Gomez-Valent:2024tdb}, the fitting value of the reduced matter density parameter in the $\Lambda$CDM, $\omega_m=\omega_b+\omega_{dm}\sim 0.139$, lies much below the value preferred by the CMB under the assumption of the standard model, $\omega_m=0.142\pm 0.001$  \cite{Planck:2018vyg}. This discrepancy is less pronounced or even non-existent in the other models. In summary, the alleviation of the Hubble tension in the $\Lambda$CDM found with angular BAO is unsatisfactory, since it comes at the expense of rocketing essentially the entire individual $\chi^2_i$ contributions, as recorded in Table \ref{tab:table_chi2}. If taken at face value, the Akaike and Deviance information criteria let us conclude that the $\CC$CDM model is, compared to the main dynamical DE models under study,  firmly ruled out when analyzed under the rich and powerful lens of the multifarious CMB+CCH+SNIa+SH0ES+BAO\_2D+$f\sigma_{12}$  observations.

It is also interesting to study what happens if we repeat the fit of the $\Lambda$CDM without including the SH0ES prior, i.e., using the data combination CMB+CCH+SNIa+BAO\_2D+$f\sigma_{12}$. We show the breakdown of the individual $\chi^2_i$ also for this case in Table \ref{tab:table_chi2}, and the fitting values of the various parameters in Table \ref{tab:table_LCDM}. As expected, the value of $\chi^2_{\rm min}$ decreases substantially compared to the case where the SH0ES prior is also considered because now the model is not penalized by its incapability of producing values of $M$ in accordance with the local SNIa calibration \cite{Riess:2021jrx}. However, we find that all the individual $\chi^2_i$ are still much larger than in $w_0w_a$CDM and $w$XCDM, which means that these models beat the $\Lambda$CDM essentially in all the data sectors, regardless of the inclusion or not of the SH0ES prior, see again Table \ref{tab:table_chi2}. They are still very strongly preferred over the $\Lambda$CDM. This remarkable result reinforces our previous arguments. The $\Lambda_s$CDM, in contrast, fails in offering an improved description of the CMB and SNIa data compared to $\Lambda$CDM when the latter is fitted without the SH0ES prior, although it still makes a better job regarding the CCH, $f\sigma_{12}$ and BAO data.

In view of the results we have obtained, we find that it is not only the dynamical character of the DE, but also the possibility that the DE changes its EoS  during different stages of the cosmic expansion, that may prove instrumental to achieve an outstanding quality fit to the overall cosmological data, compared to the $\CC$CDM. Thus, while the $w$CDM entails dynamical DE, it fails to overcome the $\CC$CDM because its EoS remains constant throughout the cosmic history. In contradistinction to that, the richer parameterization $w_0w_a$CDM with dynamical EoS performs exceptionally well and is actually also strongly preferred over the $\Lambda_s$CDM model\,\cite{Akarsu:2023mfb}, whose EoS is the same as in the $\CC$CDM. These important results were not disclosed in \cite{Gomez-Valent:2024tdb}, but now appear crystal clear in light of the values of $\Delta$AIC and $\Delta$DIC displayed in the upper half of the current Table \ref{tab:table_fits} for all models. Hence, if only from a mere quantitative point of view, speculations about a possible AdS$\to$dS transition (as first advocated in \cite{Akarsu:2023mfb}) might not be necessary if we consider the yield of the more conventional $w_0w_a$CDM description, which does not need any such exotic transition to render a much better fit than the $\CC$CDM. This is quantitatively evidenced by the substantially higher values of $\Delta$AIC ($\sim 14$ units) and $\Delta$DIC ($\sim 17$ units) obtained from $w_0w_a$CDM compared to $\Lambda_s$CDM. The large improvement of $w_0w_a$CDM over $\Lambda_s$CDM is due to a better account of the SNIa and CMB data (see the detailed breakdown of $\chi_i^2$ contributions in Table \ref{tab:table_chi2} of Appendix A).

Does this mean that transitional composite models are not competitive? Not at all, for the $w$XCDM shows no less versatility, as it has two EoS parameters to improve the description of the DE. Thus, despite the great performance of the $w_0w_a$CDM over the standard $\CC$CDM as well as over $w$CDM  and  $\CC_s$CDM, the genuine composite model $w$XCDM still beats $w_0w_a$CDM by far in all fronts, specifically with 13.9 units of $\Delta$DIC and 12.6 units of $\Delta$AIC. We refer the reader once more to Table \ref{tab:table_chi2} of Appendix A, where we can see that the $w$XCDM succeeds in decreasing the value of $\chi^2_{\rm CMB}$ by 4.4 units, the value of $\chi^2_{f\sigma_{12}}$ by 5.5 units and the value of $\chi^2_{\rm CCH}$ by 3.9 units, so the total favorable payoff comes from a rather democratic combination of background and LSS data. In particular, we may wonder why there is such a big improvement in the description of the LSS data from the $w$XCDM. Let us consider why $w_0w_a$CDM  cannot do better here.  Our constraints on the $w_0w_a$CDM parameters yield: $w_0=-0.550^{+0.084}_{-0.043}$ and $w_a=-2.04^{+0.16}_{-0.39}$. With these numbers, we can immediately check from Eq.\,\eqref{eq:CPL} that there is a crossing of the phantom divide at $z\approx 0.3$ (represented in the lower plot in Fig.,\ref{fig:background}). To be more precise, the DE exhibits phantom behavior up to $z\sim 0.3$ and
quintessence behavior for  $z\lesssim 0.3$. Since the DE behavior is phantom-like before that redshift, this compensates for the quintessence behavior after the crossing and in this way the model $w_0w_a$CDM can keep the correct distance to the last scattering surface. Hence, it is to be expected from our fit  that   $w_0w_a$CDM unduly enhances the amount of LSS compared to $w$XCDM in the late universe because there is less dark energy in the past. In fact, we obtain $\sigma_{12}=0.774$ in the best-fit $w$XCDM model, while $\sigma_{12}=0.803$ for $w_0w_a$CDM, which is obviously less competitive.

Why  $w$XCDM yields an optimal LSS description? For a better understanding  of the growth of structure in the $w$XCDM, it is worth having a closer look at the evolution of matter perturbations in this model since it is nontrivial.
In the PM phase with negative energy density and positive pressure, and for typical values of the cosmological parameters, we actually expect a larger growth of matter perturbations within the  $w$XCDM compared to the standard model and the $w_0w_a$CDM. Before the transition at $z_{t}$ (i.e. for $z>z_{t}$), the equation for the density contrast  $\delta_m=\delta\rho_m/\rho_m$ reads\,\cite{Gomez-Valent:2024tdb}
\begin{equation}\label{eq:dcX}
\delta^{\prime\prime}_m+\frac{3}{2a}\left(1-\Omega_X(a)w_X\right)\delta^\prime_m-\frac{3}{2a^2}(1-\Omega_X(a))\delta_m=0\,,
\end{equation}
with primes denoting derivatives with respect to the scale factor. PM has a negative energy density ($\Omega_X<0$) and a positive pressure $P_X=w_X\rho_X$ (owing to $w_X<-1$) and therefore induces a decrease of the friction term and an increase of the Poisson term in equation \eqref{eq:dcX}. We can define the functions

\begin{figure}[t!]
    \centering
    \includegraphics[scale=0.5]{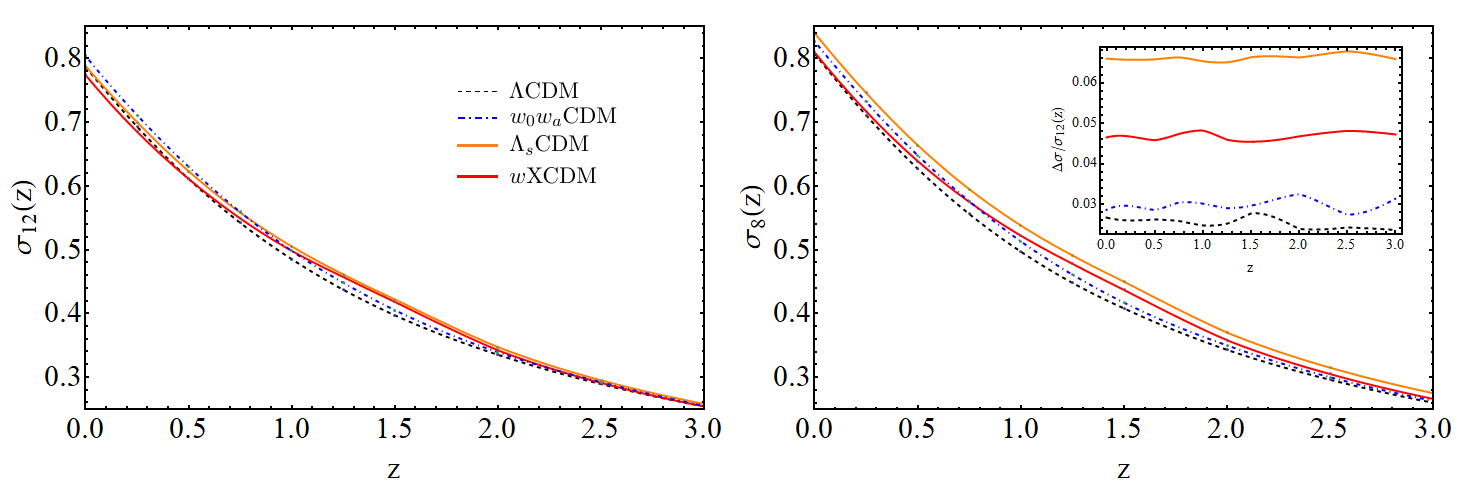} \caption{\scriptsize Comparison of the functions $\sigma_{12}(z)$ and $\sigma_8(z)$ of the various models obtained using the best-fit values from the fitting analysis with CMB+CCH+LSS+DESY5+SH0ES+BAO\_2D data (cf. Table \ref{tab:table_fits}). In the inner plot we show the relative difference $\Delta\sigma/\sigma_{12}(z)\equiv (\sigma_8(z)-\sigma_{12}(z))/\sigma_{12}(z)$.     The curves of $\sigma_8(z)$ exhibit an artificial enhancement across all redshifts, which is more pronounced for larger values of $H_0$ since $\sigma_8$ depends on the $h^{-1}$ Mpc units. This introduces an important bias that can be corrected by making use of $\sigma_{12}(z)$ instead of $\sigma_8(z)$, see \cite{Sanchez:2020vvb,Forconi:2025} for dedicated discussions on these matters.}
    \label{fig:sigmas}
\end{figure}

\begin{equation}\label{eq:F}
F(z)\equiv\Omega_{\rm DE}(z)w_{\rm DE}(z)-\Omega_{\Lambda}(z)w_{\Lambda}=\Omega_{\rm DE}(z)w_{\rm DE}(z)+\Omega_{\Lambda}(z)
\end{equation}
and
\begin{equation}\label{eq:G}
    P(z)\equiv -\Omega_{\rm DE}(z)+\Omega_\Lambda(z)
\end{equation}
in order to study the evolution of the friction and Poisson terms, respectively, entering Eq. \eqref{eq:dcX}. These two functions allow us to compare the behavior of $w$XCDM (for which $\Omega_{\rm DE}w_{\rm DE}$ is $\Omega_{X}w_X>0$ for $z>z_t$ but $\Omega_Yw_Y<0$ for $z<z_t$ ) with respect to $\Lambda$CDM ($\Omega_\CC>0$). If $F(z)$ and $P(z)$ are both positive, they favor an enhancement of matter linear perturbations in the $w$XCDM (since the latter has then less friction and less DE repulsion than  $\CC$CDM); if, on the contrary, these functions are negative, they favor a slower growth rate for $w$XCDM. We plot these functions in Fig. \ref{fig:frictionPoisson} using the best-fit values of $w$XCDM and $\Lambda$CDM displayed in the upper half of Table \ref{tab:table_fits}, obtained by using BAO 2D. As expected, they are positive before the transition, so in the presence of phantom matter there is more structure formation. This would not occur for ordinary phantom DE, for which the friction term gets enhanced and the Poisson term suppressed, i.e. just opposite to PM. Due to the phantom nature of the $X$ component, its (negative) energy density gets very diluted in the remote past compared to non-relativistic matter and eventually becomes negligible, so that these enhancement effects are not important if we go to sufficiently high redshifts\footnote{There is nonetheless a caveat here, as there could be, in principle, more PM bubbles at higher redshifts. This was noted in\,\cite{Gomez-Valent:2024tdb}. If so, they could play an important role to explain the JWST observations of extremely massive galaxies at $z\gtrsim10$\cite{Menci:2024rbq,Menci:2024hop}. The study of this possibility, however, will be presented elsewhere.}. However, these same effects enhance the power in the redshift range $z\in (1-2)$, as is apparent from the plot on the left of Fig. \ref{fig:sigmas}. Just after the transition (for $z\lesssim z_t$), DE increases up to large positive values (cf. Fig. \ref{fig:background}) and starts to evolve (decrease) as quintessence. In that part of the cosmic expansion, the equation of the density contrast takes the following form\,\cite{Gomez-Valent:2024tdb}:

\begin{figure}[t!]
    \centering
    \includegraphics[scale=0.7]{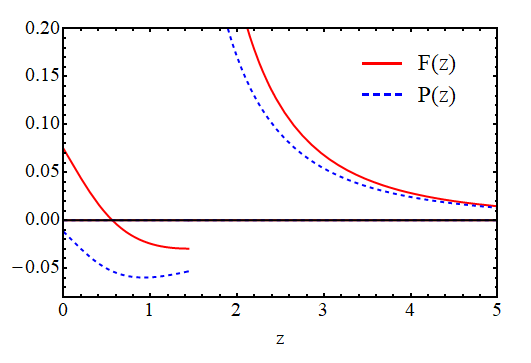} \caption{\scriptsize The functions $F(z)$ \eqref{eq:F} and $P(z)$ \eqref{eq:G} allow us to study the effect of the friction and Poisson terms appearing in the equations of the density contrast of $w$XCDM \eqref{eq:dcX}-\eqref{eq:dcY} compared to $\Lambda$CDM. Positive values of these functions indicate that the aforesaid terms favor a larger growth for $w$XCDM than for $\Lambda$CDM. This is actually what happens during the PM phase, before the transition. After it, $\Omega_{\rm DE}=\Omega_Y>\Omega_\Lambda$ (see the left upper plot of Fig. \ref{fig:background}), so the Poisson term   is smaller in the  $w$XCDM than in the $\CC$CDM. This helps to slow down the aggregation of matter in the quintessence phase of $w$XCDM. The friction term, instead, transits from negative to positive values at $z\sim 0.5$, once the deceleration parameter goes below the one of $\Lambda$CDM in magnitude, see the right upper plot in Fig. \ref{fig:background}. However, the net effect in the $w$XCDM is a decrease of the amplitude of matter fluctuations in the late universe, as it is apparent from the left plot of Fig. \ref{fig:sigmas}.}.
    \label{fig:frictionPoisson}
\end{figure}

\begin{equation}\label{eq:dcY}
\delta^{\prime\prime}_m+\frac{3}{2a}\left(1-\Omega_Y(a)w_Y\right)\delta^\prime_m-\frac{3}{2a^2}(1-\Omega_Y(a))\delta_m=0\,,
\end{equation}
with $\Omega_Y>0$ and $w_Y>-1$. The functions $F(z)$ \eqref{eq:F} and $P(z)$ \eqref{eq:G} are mostly negative  for $z<z_t$ and this fact induces a net suppression of growth that eventually brings the curve of $\sigma_{12}(z)$ of the  $w$XCDM below those of the other models in that range, and this despite the fact that the friction term becomes positive again for $z\lesssim 0.6$ (this was a non-trivial feature to be noted here) -- see again the left plot of Fig. \ref{fig:sigmas} and Fig. \ref{fig:frictionPoisson}. It is apparent that   the LSS phenomenology in the $w$XCDM model is very rich. In the remote past it can make clustering more efficient, whereas at small redshifts there can be an important suppression of growth.

In Fig. \ref{fig:sigmas}, for completeness, we also compare the evolution of the functions $\sigma_{12}(z)$ (in the left plot) and $\sigma_8(z)$ (in the right plot). This comparison is very useful to understand our choice to use $\sigma_{12}(z)$ instead of $\sigma_{8}(z)$,  following our previous work \cite{Gomez-Valent:2024tdb}. The common practice of using $\sigma_8(z)$ introduces a bias that becomes more significant for larger values of $H_0$ owing to the dependence of the scale $R_8$ on $h$, specifically on the $h^{-1}$ Mpc units \cite{Sanchez:2020vvb,Forconi:2025}. This bias induces an artificial relative shift that is approximately independent of the redshift \cite{Forconi:2025}. As a result, it can even invert the proper classification of models based on their ability to explain the LSS of the universe \cite{Forconi:2025}. Notice in Fig. \ref{fig:sigmas} that the curves of $\sigma_{12}(z)$ for the various models seem to converge at high redshifts, meaning that the data are able to constrain the amplitude of fluctuations at linear scales for all models in a consistent way. In transitional models, this is possible despite the presence of PM because the best-fit values of the amplitude of the power spectrum are much lower than those encountered in the $\Lambda$CDM and the $w_0w_a$CDM. The convergence of $\sigma_{12}(z)$ at high redshifts is a feature that is also lost if we analyze the results in terms of $\sigma_{8}(z)$. In the inner plot of Fig. \ref{fig:sigmas} we show the bias introduced by $\sigma_8(z)$, obtained using our best-fit models from the fitting analyses with CMB+CCH+LSS+DESY5+SH0ES+BAO\_2D data (cf. Table \ref{tab:table_fits}). The use of $\sigma_{12}(z)$ eliminates the aforementioned bias because it characterizes the amplitude of the power spectrum at a fixed scale $R_{12}=12$ Mpc \cite{Sanchez:2020vvb}, and this is why we adopt it in our work.

The results we have obtained with the data set CMB+CCH+DESY5+SH0ES+BAO\_2D+$f\sigma_{12}$ are fully consistent with those presented in \cite{Gomez-Valent:2024tdb}.  However, replacement of Pantheon+ SNIa data with DES-Y5 data enhances the signal in favor of $w$XCDM compared to $\Lambda_s$CDM and allows excluding $\Lambda$CDM with still higher statistical significance. Moreover, we have checked that the $w_0w_a$CDM parameterization is also able to fit very efficiently this data set, in actual fact improving the performance of $\Lambda_s$CDM, despite providing a weaker fit to the LSS data. To the best of our knowledge, the present work  is the first time in the literature wherein the $w_0w_a$CDM parameterization is constrained using BAO 2D measurements.

We turn now the discussion to the results of our analysis using BAO 3D-only, in combination with the rest of the data sources. Due to the existing tension between the angular and anisotropic BAO data, and given the pivotal role that BAO plays in the building of the inverse distance ladder and the discussion of the Hubble tension \cite{Camarena:2019rmj,Gomez-Valent:2023uof,Favale:2024sdq}, it is of utmost importance to study how the results of our analysis change when we replace  BAO 2D data with BAO 3D data. This represents one of the primary objectives of this paper. The fitting results obtained with CMB+CCH+DESY5+SH0ES+BAO\_3D+$f\sigma_{12}$ data are displayed in the lower half of Table \ref{tab:table_fits}. We observe that there is again very strong evidence against $\Lambda$CDM and in favor of the other models (except once more for the $w$CDM), since we find $\Delta$DIC$>10$ in all relevant cases, accompanied with  consistent values of $\Delta$AIC. Notwithstanding, it is a fact that BAO 3D does not give room for a notable increase of $H_0$. But this should not come as a big surprise, for we know that in order to solve the Hubble tension,  BAO 3D does not only require an abrupt (phantom-like) increase of $\rho_{\rm DE}$ at very small redshifts ($z\lesssim 0.15-0.2$), but also a simultaneous increase of the absolute magnitude of SNIa \cite{Alestas:2020zol,Alestas:2021luu,Gomez-Valent:2023uof}, a scenario  that we do not contemplate in this work. The tension, when analyzed in terms of $M$ (which is the primary measurement of SH0ES) remains at $4.32\sigma$ ($\Lambda$CDM), $4.46\sigma$ ($w$CDM),  $3.74\sigma$ ($w_0w_a$CDM), $3.60\sigma$ ($w$XCDM) and $3.44\sigma$ ($\Lambda_s$CDM), so the tension is quite high regardless of the model, this being the case even including the SH0ES prior in the fitting analysis, cf. Sec. \ref{sec:Data}. Instead, if we quantify the tension in terms of the $H_0$ values, we obtain the following results: $1.7\sigma$ ($\Lambda$CDM), $2.0\sigma$ ($w$CDM), $2.4\sigma$ ($w_0w_a$CDM), $1.7\sigma$ ($w$XCDM) and $\sim 1\sigma$ ($\Lambda_s$CDM)\footnote{It has been argued in many papers that the true tension has to be quantified in terms of $M$, see, e.g., \cite{Camarena:2019moy,Benevento:2020fev,Efstathiou:2021ocp}. Notice that the Hubble tension evaluated in terms of $H_0$ and using the distance ladder measurement of the Hubble function obtained from the SH0ES calibration of $M$ and the SNIa in the Hubble flow from DES-Y5 ($H_0=70.5\pm1.1$ km/s/Mpc \cite{Favale:2025}) is found to be artificially small when BAO 3D is employed in the fitting analyses.}. All these features were already advanced in our conclusions of \cite{Gomez-Valent:2024tdb}, but now we can reply the question of why the $\Lambda$CDM is also strongly disfavored in light of the dataset CMB+CCH+SNIa+SH0ES+BAO\_3D+$f\sigma_{12}$. The answer is given in the breakdown of individual $\chi_i^2$ contributions displayed in the  lower half of Table \ref{tab:table_chi2}. On the one hand, the transitional DE models and the $w_0w_a$CDM fit better both the CMB and DESY5+SH0ES data sets, offering a substantial decrease of the corresponding $\chi^2_i$ yields. But on the other, the $w_0w_a$CDM is now able to improve significantly the description 
of  BAO 3D compared to the transitional models. This is in contradistinction to the situation found before with BAO 2D, and this explains why the $w_0w_a$CDM parameterization leads to the largest values of $\Delta$DIC and $\Delta$AIC in the presence of BAO 3D.  Even so the difference with $w$XCDM is not significant.  For instance, $\Delta$DIC is 1.6 and 6.8 units larger for $w_0w_a$CDM than for $w$XCDM and $\Lambda_s$CDM, respectively. The fitting values of the $w_0w_a$CDM parameters read, in this case: $w_0=-0.762\pm 0.039$ and $w_a=-0.87^{+0.14}_{-0.16}$, which are in full accordance with those reported by DESI \cite{DESI:2024mwx,DESI:2024hhd} and other works that, like us, do not include DESI data \cite{Chan-GyungPark:2024spk,Chan-GyungPark:2024brx}. For instance, using DESI(FS+BAO)+CMB+DESY5 the authors of \cite{DESI:2024hhd} find $w_0=-0.761\pm 0.065$ and $w_a=-0.96^{+0.30}_{-0.26}$, whereas using DESI(FS+BAO)+CMB+PantheonPlus, $w_0=-0.858\pm 0.061$ and $w_a=-0.68^{+0.27}_{-0.23}$. The agreement with our results is evident, although our uncertainties are smaller, also for $\Omega_m^0$ and $H_0$ (cf. our Table \ref{tab:table_fits} and Table 2 of \cite{DESI:2024hhd}). For $w_0$ and $w_a$ we find a significant decrease in the errors by a factor of $1.6$–$1.7$. Using the best-fit values of these parameters we find that the crossing of the phantom divide now occurs at $z\sim 0.4$, quite close to the one found with BAO 2D.

Our constraints at $68\%$ CL on $w_X$ and $w_Y$ in $w$XCDM now read: $w_X<-1.08$ and $w_Y=-0.933\pm
 0.025$. They are consistent with the results obtained with BAO 2D and, again,  favor quintessence evolution of the DE after the transition redshift and a PM phase with $w_X<-1$ before the transition. However, it is important to comment on the fact that we find $w_X<-0.75$ at $95\%$ CL, so the dataset built with BAO 3D does not exclude the region $w_X>-1$ in a statistically significant way. The transition redshift is moved from $z_t\sim 1.5$ (with BAO 2D) to $z_t\sim 2$ (with BAO 3D), hence it lies almost in the region where there is no background nor LSS data. In point of fact, we have checked that our data set does not exclude the possibility of this transition to occur at much larger redshifts if $w_X$ approaches 0 from below with increasing accuracy, since this is sufficient to produce the correct distance to the last scattering surface. These effects complicate the convergence of the chain. We have imposed a prior $w_X\in [-2,-0.5]$ to keep them under control. The data set CMB+CCH+DESY5+SH0ES+BAO\_3D+$f\sigma_{12}$ sets strong constraints on $w_Y$, but the actual constraints on $z_t$ and $w_X$ would be quite loose in the absence of this prior.

 The fit to the LSS data is very similar in all the models under study when we consider BAO 3D. We find values of $\sigma_{12}\sim 0.79-0.80$ for all of them. We deem important also to highlight some other aspects that can be found when looking carefully at Table \ref{tab:table_chi2}: (i) The values of $\chi^2_{{\rm DESY5+SH0ES}}$ obtained in the analyses with BAO 3D are significantly larger in the $w_0w_a$CDM and $w$XCDM models compared to those obtained in the analyses with BAO 2D. This is essentially due to the Hubble tension, which is not cured if we use anisotropic BAO. In contast, for $\Lambda_s$CDM the value remains quite stable and larger than in the other non-standard cosmologies in both analyses. This might be related to the fact that DES-Y5 data prefers dynamical DE rather than a rigid $\CC$ term (even if changing sign at a transition point); (ii) CMB data is much better accomodated by $\Lambda$CDM if use is made of BAO 3D. However, the other models still decrease  $\chi^2_{\rm CMB}$ by $\sim 3$ units.

To summarize, although we find a very different response of the various models studied in this work to the cosmological tensions when we build the inverse distance ladder using angular or anisotropic BAO data, we consistently find evidence supporting physics beyond $\Lambda$CDM, regardless of the BAO type and using a very rich and diverse set of cosmological measurements. This fact alone is truly remarkable from our point of view and is one of the main conclusions of this work.


\section{Conclusions}\label{sec:Discussion}
In this Letter, we have assessed the impact of composite DE scenarios versus monocomponent scenarios concerning  a possible resolution of the cosmological tensions\cite{Abdalla:2022yfr} We have revisited standard parameterizations of the DE, such as $w$CDM\cite{Turner:1997npq} and  $w_0w_a$CDM \cite{Chevallier:2000qy,Linder:2002et} assuming one single DE fluid against transitional models with two fluids, wherein the DE density undergoes a sign flip at some point in the `recent' past (typically at around $z_t\simeq 1-2$). We have compared the results of these parameterizations with the model $w$XCDM\,\cite{Gomez-Valent:2024tdb},  which we use  as an archetype of composite model exhibiting an EoS transition at a point $z_t$. e have also confronted all these models with the simplest situation, in which the DE is represented by a rigid cosmological term, i.e. the standard $\CC$CDM model. We have found that the main dynamical DE models studied here (above all $w$XCDM and  $w_0w_a$CDM) prove to be much more efficient than the $\CC$CDM to fit the overall cosmological data. This is a major result of our study, which is fully in line with previous large scale analyses of cosmological data made a few years ago\,\cite{Sola:2015wwa,Sola:2016jky,SolaPeracaula:2016qlq,SolaPeracaula:2017esw,Sahni:2014ooa,Gomez-Valent:2014rxa,Zhao:2017cud,SolaPeracaula:2018wwm}, in which dynamical DE was already being strongly advocated.  In particular, the suitability of the dynamical character of the DE also applies to the composite models under discussion; and we should furthermore stress that the statement holds good both under transversal-only and anisotropic-only BAO data (or BAO 2D and BAO 3D, respectively, for short). This result is highly remarkable in itself, especially if we take into account that these two versions of the BAO data are still in conflict; see, in particular, the analyses of \cite{Favale:2024sdq,Camarena:2019rmj,Gomez-Valent:2023uof} and references therein. We cannot exclude the fact that the ultimate reason (or at least a major influence) behind the $H_0$-tension lies precisely in such a BAO conflict. But even if it is the case, we have proven here that the overall fit with all the cosmological data, which means that with any of these BAO types alone together with the remaining CMB+CCH+LSS+DESY5+SH0ES data sets, can be significantly improved with dynamical models of the DE and particularly with composite DE models involving a sign flip of the energy density at a transition redshift $z_t$.  We have also shown that exceedingly simple DE parameterizations with a fixed EoS parameter, such as the $w$CDM, fall short to improve the yield of the standard $\CC$CDM.  Successful parameterizations require dynamical EoS or a composite DE structure with more than one EoS parameter acting at different stages of the cosmic evolution.

As indicated, the basic composite model that we have used to illustrate our claim is the $w$XCDM model proposed in \cite{Gomez-Valent:2024tdb}, consisting of two DE components $X$ and $Y$ acting in sequence and possessing respective EoS parameters $w_X$ (for $z>z_t$) and $w_Y$ (for $z<z_t$) together with sign-flipped DE densities $\rho_X<0$ and $\rho_Y>0$. We have compared this model with a particular implementation of it (or at least it can be viewed effectively as such), the so-called  $\CC_s$CDM model\cite{Akarsu:2023mfb,Akarsu:2021fol}, in which the two EoS parameters $w_X$ and $w_Y$ are both fixed at $-1$ (corresponding to a cosmological constant $\CC$) with a sign flip of $\CC$ at the transition point $z_t$. Despite this setting reducing the number of parameters to just one ($z_t$), our analysis clearly demonstrates that if  $w_X$  and $w_Y$ are left as part of the fitting degrees of freedom (as is indeed the case in the $w$XCDM) the increase in quality of the fit can be so high that it more than compensates for the penalty received by the information criteria for the two extra free parameters that the $w$XCDM has as compared to the $\CC_s$CDM. The outcome is that $w$XCDM performs optimally and, in addition, it provides dynamical DE results compatible with the last DESI results\,\cite{DESI:2024mwx,DESI:2024hhd} since the late time component of $w$XCDM, i.e. $Y$,  is found to have quintessence-like EoS behavior ($w_Y\gtrsim-1$). Obviously, this feature is not possible within the rigid structure of  the $\CC_s$CDM.  We also note that despite the fact that the $\CC_s$CDM model performs better than the $\CC$CDM (as a global fit), it nevertheless performs less than the $w_0w_a$CDM parameterization under the same data. Furthermore, we understand that the composite scenarios considered here may actually serve as generic parameterizations for possible fundamental models of the DE. For example, as previously pointed out, the $w$XCDM composite structure with phantom matter (PM) mimics the old $\CC$XCDM model\cite{Grande:2006nn} and it also appears in stringy versions of the running vacuum model (RVM)\cite{Mavromatos:2021urx,Mavromatos:2020kzj}.

Although  the overall fit quality obtained from the composite DE scenarios can be much better than the $\CC$CDM,  we find that insofar as concerns the specific amelioration of the tensions, the result depends on whether our global fit is performed using  BAO 2D (only) or BAO 3D (only). Remarkably, the growth tension can be highly relieved for all models when expressed in terms of the $\sigma_{12}$ parameter\,\cite{Sanchez:2020vvb}, the alleviation being complete only in the $w$XCDM under BAO 2D. However, the $H_0$ tension can be fully cut down only in the BAO 2D case, but not for BAO 3D.

All in all, the dynamical DE models admitting a variable EoS and/or a transitional EoS structure around a  redshift $z_t\simeq 1-2$ prove highly successful in improving the standard model description of the cosmological data. In particular, the archetype composite model $w$XCDM with a transition redshift near $z_t\simeq 1.4$  offers an outstanding quality fit with BAO 2D data and a possible  solution to the cosmological tensions.  In addition, this model mimics the conceptual framework capable of affording a better understanding of the potentially involved physical phenomena, namely the stringy RVM\,\cite{Mavromatos:2021urx,Mavromatos:2020kzj}.   As explained in \cite{Gomez-Valent:2024tdb}, within this fundamental framework the generation of a phantom matter (PM) bubble is induced by quantum fluctuations. The phenomenon can be iterated throughout the cosmic expansion:  bubbles of PM  might lurk behind the anomalous outgrowth of structures found at redshifts in the high range  $z\sim 10$. If so, the $w$XCDM scenario could offer a possible explanation for the appearance of the supermassive galaxies recently spotted by the JWST mission\,\cite{Labbe:2022ahb,Adil:2023ara,Menci:2024rbq}. We believe that such a possibility deserves a separate study.

\vspace{0.3cm}
{\bf Acknowledgments}
This work is  partially supported by grants PID2022-136224NB-C21 and  PID2019-105614GB-C21, from MCIN/AEI/10.13039/501100011033.  AGV is funded by “la Caixa” Foundation (ID 100010434) and the European Union's Horizon 2020 research and innovation programme under the Marie Sklodowska-Curie grant agreement No 847648, with fellowship code LCF/BQ/PI23/11970027. JSP is funded also  by  2021-SGR-249 (Generalitat de Catalunya) and
CEX2019-000918-M (ICCUB, Barcelona).  Both of us  acknowledge networking support by the COST Association Action CA21136 ``{\it Addressing observational tensions in cosmology
with systematics and fundamental physics (CosmoVerse)}''.
\color{black}

\newpage
\appendix
\section*{Appendix A: Breakdown of $\chi^2_{\rm min}$ contributions}

\begin{table}[ht!]
\centering
\renewcommand{\arraystretch}{1.4}
\begin{tabular}{|c ||c |c |c|c |c |}
\hline
\multicolumn{6}{|c|}{\textbf{CMB+CCH+SNIa+SH0ES+BAO\_2D+$f\sigma_{12}$}}\\\hline\hline\hline
{\small $\chi^2_i$} & {\small $\Lambda$CDM} & {\small $w$CDM}   & {\small $w_0w_a$CDM } & {\small $w$XCDM }  & {\small $\Lambda_s$CDM}
\\\hline
$\chi^2_{\rm Planck\_highl\_TTTEEE}$ & 2359.51 (2355.29) & 2356.96 & 2353.48 & 2350.12 & 2352.81 \\\hline
$\chi^2_{\rm Planck\_lowl\_EE}$ & 400.40 (397.56) & 396.10 & 397.30 & 395.89 & 396.43 \\\hline
$\chi^2_{\rm Planck\_lowl\_TT}$ &  22.07 (22.47) & 22.17 & 21.65 & 25.80 & 28.00 \\\hline
$\chi^2_{\rm Planck\_lens}$ &  10.45 (10.01)  & 10.47 & 8.45 & 8.59 & 11.47 \\\hline\hline
$\chi^2_{\rm CMB}$ & 2792.43 (2785.33) & 2785.70 & 2780.88 & 2780.40 & 2788.71 \\\hline
$\chi^2_{{\rm DESY5+SH0ES}}$ & 1671.11 (1650.67) & 1674.57 & 1646.11 & 1641.71 & 1666.59  \\\hline
$\chi^2_{\rm BAO}$ & 29.33 (35.22) & 30.90 & 14.18 & 13.95 & 14.24 \\\hline
$\chi^2_{f\sigma_{12}}$ & 17.88 (20.03) & 18.93 & 17.56 & 12.04 & 9.34 \\\hline
$\chi^2_{\rm CCH}$ & 12.88 (13.07) & 12.9 & 13.68 & 9.75 & 9.70 \\\hline\hline
$\chi^2_{\rm min}$ & $4523.64$ (4504.32) & 4523.60 & 4472.42 & $4457.84$ &  $4488.56$\\\hline\hline\hline
\multicolumn{6}{|c|}{\textbf{CMB+CCH+SNIa+SH0ES+BAO\_3D+$f\sigma_{12}$}} \\
\hline\hline
\hline
{\small $\chi^2_i$} & {\small $\Lambda$CDM} & {\small $w$CDM}   & {\small $w_0w_a$CDM } & {\small $w$XCDM }  & {\small $\Lambda_s$CDM}
\\\hline
$\chi^2_{\rm Planck\_highl\_TTTEEE}$ & 2354.72 & 2356.24 & 2352.48 & 2351.93 & 2352.27 \\\hline
$\chi^2_{\rm Planck\_lowl\_EE}$ & 396.79 & 399.63 & 396.67 & 395.87 & 395.71 \\\hline
$\chi^2_{\rm Planck\_lowl\_TT}$ & 22.93 & 21.75 & 23.50 & 23.09  & 23.83 \\\hline
$\chi^2_{\rm Planck\_lens}$ &  9.43 & 9.81 & 8.75 & 8.99 & 8.79 \\\hline\hline
$\chi^2_{\rm CMB}$ & 2783.87 & 2787.43 & 2780.40 & 2779.88 & 2780.60 \\\hline
$\chi^2_{{\rm DESY5+SH0ES}}$ & 1672.52 & 1670.64 & 1656.81 & 1656.32&  1664.51  \\\hline
$\chi^2_{\rm BAO}$ &  29.44 & 27.75 & 26.77 & 33.19&  33.97 \\\hline
$\chi^2_{f\sigma_{12}}$ & 15.86 & 15.20 & 16.98 & 15.16 & 13.22 \\\hline
$\chi^2_{\rm CCH}$ & 13.13 & 13.16 & 12.97 & 12.91 & 12.60 \\\hline\hline
$\chi^2_{\rm min}$ & 4514.84 & 4514.20 & 4493.92 & 4497.46  & 4504.92 \\\hline
\end{tabular}
\caption{\scriptsize Individual $\chi^2_i$ contributing to $\chi^2_{\rm min}$, obtained in the fitting analyses for the various models with CMB+CCH+SNIa+SH0ES+BAO\_2D+$f\sigma_{12}$ (upper half) and CMB+CCH+SNIa+SH0ES+BAO\_3D+$f\sigma_{12}$ (lower half). $\chi^2_{\rm CMB}$ contains the total CMB contribution, i.e. it is the sum of all the Planck $\chi^2_i$. For the analysis of the $\Lambda$CDM with BAO 2D, we also show in brackets the results obtained without using the SH0ES prior.}
\label{tab:table_chi2}
\end{table}

\newpage

\section*{Appendix B: Fitting results for the $\Lambda$CDM with and without the SH0ES prior}

\begin{table}[h!]
\centering
\renewcommand{\arraystretch}{1.74}
\resizebox{\textwidth}{!}{
\begin{tabular}{|c ||c |c | }
\hline
& CMB+CCH+SNIa+SH0ES+BAO\_2D+$f\sigma_{12}$ & CMB+CCH+SNIa+BAO\_2D+$f\sigma_{12}$ \\
\hline
\hline
\hline
{\small Parameter} & {\small $\Lambda$CDM} & {\small $\Lambda$CDM }
\\\hline
$10^2\omega_b$ & $2.271\pm 0.014$ (2.267) & $2.256\pm 0.015 $ (2.257)   \\\hline
$10\,\omega_{\rm dm}$ &  $1.160\pm 0.009$ (1.159)& $1.173\pm 0.009$ (1.174)  \\\hline
$\ln(10^{10}A_s)$ & $3.060^{+0.016}_{-0.018}$ (3.067) & $3.054\pm 0.016$ (3.050)  \\\hline
$\tau$ & $0.066^{+0.008}_{-0.009}$ (0.068) & $0.062^{+0.008}_{-0.009}$ (0.061)   \\\hline
$n_{s}$ &  $0.975\pm 0.004$ (0.974) & $0.972\pm 0.004$ (0.971)  \\\hline
$H_{0}$  &  $69.28\pm 0.41$ (69.22) & $68.61\pm 0.43$ (68.56)   \\\hline\hline
$\Omega_m^0$ & $0.290\pm 0.005$ (0.291) & $0.299\pm 0.005$ (0.299)  \\\hline
$M$ &  $-19.360\pm 0.010$ (-19.362) & $-19.378\pm 0.011$ (-19.380)  \\\hline
$\sigma_{12}$ & $0.784\pm 0.007$ (0.787) & $0.791\pm 0.006$ (0.790)
 \\\hline\hline
$\chi^2_{\rm min}$ &  $4523.64$ & $4504.32$  \\\hline\hline
\end{tabular}}
\caption{\scriptsize Mean values and uncertainties at 68\% CL obtained for the $\Lambda$CDM with the data set CMB+CCH+SNIa+BAO\_2D+$f\sigma_{12}$ with and without the inclusion of the $H_0$ prior.}
\label{tab:table_LCDM}
\end{table}

%
\clearpage


\end{document}